# Wolf phase tomography (WPT)

# of transparent structures using partially coherent illumination

*Xi Chen, Mikhail E. Kandel, Chenfei Hu, Young Jae Lee, and Gabriel Popescu*

Quantitative Light Imaging Laboratory, Beckman Institute for Advanced Science and Technology, Department of Electrical and Computer Engineering, University of Illinois at Urbana-Champaign, Urbana, Illinois 61801, USA

The official email addresses of all authors:

Xi Chen: xic@illinois.edu

Mikhail E. Kandel: kandel3@illinois.edu

Chenfei Hu: chenfei3@illinois.edu

Young Jae Lee: lee134@illinois.edu

Gabriel Popescu: gpopescu@illinois.edu

Corresponding author:

Gabriel Popescu, 4055 Beckman Institute, 405 North Mathews Ave, Urbana, Illinois 61801, (217) 333-4840, gpopescu@illinois.edu

**Abstract**

Diffraction tomography using coherent holographic imaging has been proposed by Emil Wolf in 1969 to extract 3D information from transparent, inhomogeneous objects. At the same time, the Wolf equations describe the propagation correlations associated with partially coherent fields. Combining these two concepts, here we present Wolf phase tomography (WPT), which is a method for performing diffraction tomography using partially coherent fields. The WPT reconstruction works in the direct space-time domain, without the need of Fourier transformation, and decouples the refractive index distribution from the thickness of the sample. We demonstrate the WPT principle using data acquired by spatial light interference microscopy (SLIM). SLIM is a quantitative phase imaging method that upgrades an existing phase contrast microscope by introducing controlled phase shifts between the incident and scattered fields. The illumination field in SLIM is spatially partially coherent (emerging from a ring-shaped pupil function) and of low temporal coherence (white light), thus, suitable for the Wolf equations. From three intensity measurements corresponding to different phase-contrast frames in SLIM, the 3D refractive index distribution is obtained right away by computing the Laplacian and second time derivative of the measured complex correlation function. The high-throughput and simplicity of this method enables the study of 3D, dynamic events in living cells across entire multi-well plates, with RI sensitivity on the order of $10^{-5}$. We validate WPT with measurements of standard samples (microbeads), spermatozoa, and live neural networks.

# 1 Introduction

Refractive index (RI) is a fundamental physical property that determines how light interacts with a medium in terms of *scattering*, governed by its real part, and *absorption*, through its imaginary part [1-5]. In biological applications, the RI distribution highly correlates with cellular properties, such as dry mass and chemical concentrations [6-9]. Tissue RI can also act as an intrinsic marker for cancer diagnosis [10, 11]. Nanoscale morphological changes in cells and tissues can be revealed by the RI maps [12, 13]. For example, it has been shown that cancer tissue exhibits higher RI variances than normal tissue [10, 11]. RI can also be used to study biological events including cellular transport and mitosis [14, 15], and can be used for phenotypic screening and cellular monitoring [16, 17]. In order to obtain the RI distribution of cells and tissues from the measured field properties in different imaging modalities, one must go beyond the typical quantities measured in phase imaging and solve a scattering inverse problem. A condition for this problem to yield unique solutions is to measure the full information about the scattered field, meaning, both the amplitude and phase. Interferometric microscopy provides a method for phase retrieval of weak scattered samples such as cells and tissues [18-20].

Quantitative phase imaging (QPI) has emerged as a dynamic field focused entirely on extracting phase distributions of an imaging field and exploiting that information for biomedical applications [18, 21-27]. White light-based methods, such as SLIM [28], gradient light interferometric microscopy GLIM [29], and (white-light) diffraction phase microscopy (DPM) [30] can render phase images of live cells without the speckle noise typically associated with coherent illumination [21]. As a result, the spatial sensitivity to pathlength changes is very high. The optical pathlength measurement depends on both the RI and the thickness of the sample [31]. In order to estimate the 2D (axially-averaged) RI from the optical pathlength, the thickness distribution of the structures needs to be known or decoupled from the optical pathlength [32, 33]. However, the accuracy is low due to the geometrical optics approximation and the results provide only a 2D map of the longitudinally averaged RI.

For inferring the 3D RI distribution from the QPI data, several approaches have been proposed based on solving the deterministic wave equations [34]. One of them is the filtered back-projection algorithm, using the Fourier diffraction theorem and first-order Born or Rytov approximation [25]. It connects the object function with the Fourier transform of the projection. The reconstruction of the RI distribution is obtained by combining the frequency bands with respect to different angles [35]. One can achieve this by rotating the illumination angels or measuring a set of image fields at successive points across a cell with a focused beam illumination, also known as synthetic aperture tomography phase microscopy [36]. However, the Fourier diffraction theorem [34] assumes plane wave illumination and is only an approximation for partially coherent fields. In order to obtain more accurate solutions to the inverse problem, taking the coherence properties into account is inevitable. White-light diffraction tomography (WDT) uses the temporal correlation and instrument response to perform a deconvolution on the complex field data to extract the 3D scattering potential [37]. However, WDT requires *a priori* knowledge of the instrument impulse response (or transfer function), which is often limited. At the same time, deconvolution operations are usually time-consuming and sensitive to noise.

The transport of intensity equation can connect the RI to the intensity of bright field images [38, 39]. Nonetheless, this method is only applicable under the paraxial approximation (see, e.g., Section 12.1 in [18]). Integrated microchips have been used to measure RI information, combining an external cavity laser, micro-lenses, and microfluidic channels into a monolithic device [40, 41]. It can determine the average RI of a single live cell in real-time but cannot render the RI distribution. 3D RI distributions are usually constructed through axial scanning (z-scanning) [29, 37, 42] or projections from different angles (computed tomography) [25, 26, 43]. To increase the axial resolution in the 3D reconstruction, efforts have been devoted toward alleviating the incomplete frequency coverage of any imaging system, or "the missing cone problem" [44]. Illumination angle-scanning and rotation of the sample can help fill in the missing cone region. Cells can be rotated by optical tweezers or dielectrophoretic forces in microfluidics [45-48]. However, these methods involve more complicated procedures.

In this paper, we proposed a fast 3D RI construction method, based on the Wolf equations for propagating correlations of partially coherent light [49, 50]. This approach, referred to as Wolf phase tomography (WPT), involves minimal computational steps, renders high-resolution RI tomograms, without time-consuming deconvolution operations. WPT decouples the refractive index distribution from the thickness of the sample directly in the space-time domain, without the need for Fourier transformation. We demonstrate that, from three independent intensity measurements corresponding to each phase shift in SLIM, the RI distribution is reconstructed right away from the Laplacian and second time derivative of the complex correlation functions. We demonstrate the WPT with standard polystyrene beads, fixed spermatozoa, and dynamic live cell imaging over many hours. Interestingly, we found that WPT is able to extract intrinsic refractive index changes in live cells with a sensitivity on the order of $10^{-5}$, which can report on cell viability.

## 2 Results

**WPT principle.** WPT relies on a commercial phase contrast microscope upgraded with an add-on SLIM module (CellVista SLIM Pro, Phi Optics, Inc.), as shown in Figure. 1a. Figures 1b-d show the temporal spectrum and autocorrelation properties of the illumination (white-light) field. In addition to the $\pi/2$ phase shift between the incident and scattered fields introduced by the objective phase ring, the SLIM module provides further phase shifts with $\pi/2$ increments. At the camera plane, we record three intensity images, corresponding to each phase shift, as illustrated in Fig. 1e, namely,

$$I_d(\mathbf{r}) = I_i(\mathbf{r}) + I_s(\mathbf{r}) + 2\Re\left[\Gamma_{is}\left(\omega_0\tau_d + \Delta\phi(\mathbf{r})\right)\right], \tag{1}$$

where $\omega_0\tau_d = -d\pi/2$, $d=1,2,3$, $\omega_0$ is the central frequency of the incident field, $\Re$ stands for the real part, $\Delta\phi$ is the phase difference between the incident field, $U_i$, and scattered field, $U_s$, and $\Gamma_{pq}(\mathbf{r}_1,\mathbf{r}_2,\tau) = \langle U_p^*(\mathbf{r}_1,t)U_q(\mathbf{r}_2,t+\tau)\rangle_t$, $p,q=\{i,s\}$. From these three frames, we are able to solve for $\Re\left[\Gamma_{is}\left(\omega_0\tau_d + \Delta\phi(\mathbf{r})\right)\right]$. Based on partially coherent light propagation, governed by the Wolf equations [50], the refractive index of the object can be obtained by (see the full derivation in Supplementary Note 1)

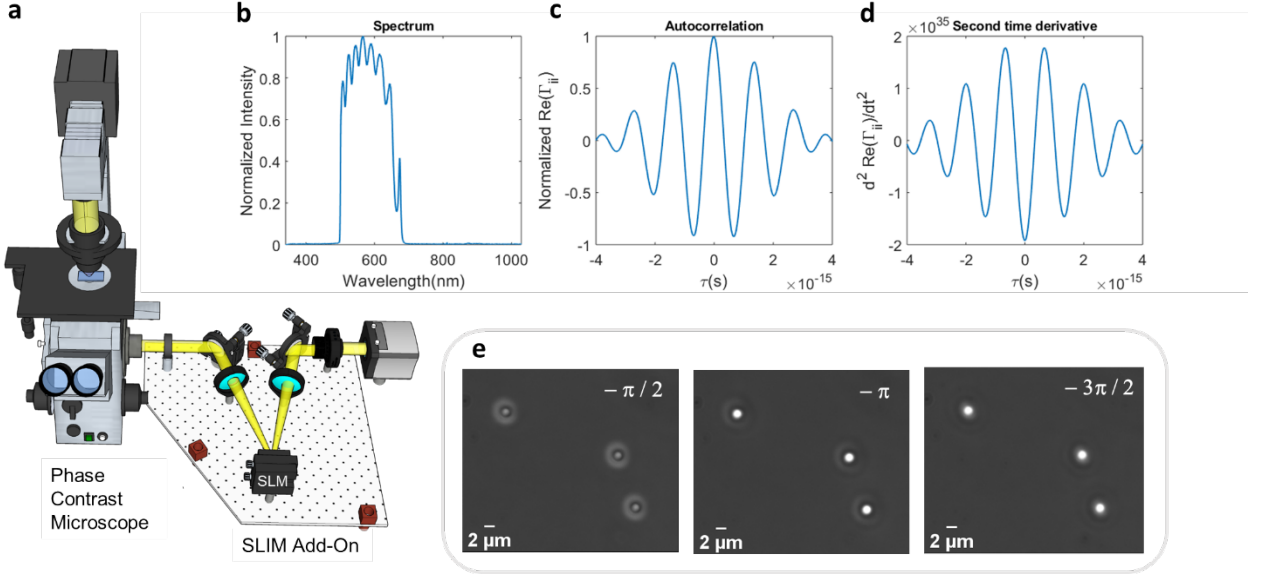

Fig. 1 **Optical setup and working principle of WPT. a** WPT optical setup. The SLIM add-on module is mounted to the output camera port of a phase contrast microscope. This module shifts the phase between the incident and scattered field with a π/2 increment using SLM. The interference patterns are recorded by the camera. **b** Spectrum, **c** Autocorrelation and **d** Second-order derivative of the autocorrelation of the halogen source measured by the spectrometer. **e** Three phase-shifted frames of 2μm polystyrene beads are acquired using the SLIM module (63x/1.4NA objective).

$$n(\mathbf{r}) = \sqrt{\frac{m(\mathbf{r}) - n_0^2 [1 - g(\mathbf{r})]}{1 + g(\mathbf{r})}}. \qquad (2)$$

In Eq. 2, the functions *m* and *g* are defined as

$$m(\mathbf{r}) = \left. \frac{c^2 \nabla^2 \Re[\Gamma_{is}(\mathbf{r},\mathbf{r},\omega_0\tau)]}{\dfrac{\partial^2 \Re[\Gamma_{is}(\mathbf{r},\mathbf{r},\omega_0\tau)]}{\partial \tau^2}} \right|_{\omega_0\tau=-\pi}, \qquad (3a)$$

$$g(\mathbf{r}) = \left. \frac{\dfrac{\partial^2 \Re[\Gamma_{ii}(\mathbf{r},\mathbf{r},\omega_0\tau)]}{\partial \tau^2}}{\dfrac{\partial^2 \Re[\Gamma_{is}(\mathbf{r},\mathbf{r},\omega_0\tau)]}{\partial \tau^2}} \right|_{\omega_0\tau=-\pi}, \qquad (3b)$$

with *r=(x, y, z)* the spatial coordinate, $n_0$ the refractive index of the background media and *c* the speed of light in vacuum. Figure 1b describes the normalized spectrum of the halogen source measured by the spectrometer (Ocean Optics). The real part of the normalized autocorrelation $\Re[\Gamma_{ii}(\mathbf{r},\mathbf{r},\omega_0\tau)]$ is obtained by taking the Fourier transform of the spectrum (see Fig. 1c). In order to retrieve the temporal correlation function quantitatively, we normalized the $\Gamma_{ii}(\mathbf{r},\mathbf{r},0)$ value from the spectrometer data to the background intensity from the camera, corrected with the spectrally dependent quantum efficiency of the camera. Thus, we ensured that the autocorrelation $\Gamma_{ii}(\mathbf{r},\mathbf{r},0)$ measured by the two different devices has the same value. The second-order time derivative of $\Gamma_{ii}(\mathbf{r},\mathbf{r},\tau)$ is depicted in Fig. 1d. The Laplacian in Eq. (3a) is calculated using three images with the first-order finite difference approximation. The z-component of the Laplacian was computed using three axially distributed frames, separated by a distance that matches the x-y pixel sampling and is much smaller than the diffraction spot. For example, for a 40X/0.75NA objective, this distance is 0.14 μm, while the diffraction-limited resolution is 0.4 μm. The second-order derivatives in Eq. (3a-3b) are calculated in MATLAB, using three phase-shifted frames. Smaller phase shifts would give more accurate derivatives. However, the contrast between different frames would largely decrease, thus the signal to noise ratio would decrease as well, resulting in lower accuracy of the derivative. Therefore, in order to increase the signal to noise ratio and accuracy, we keep the phase increment at $\pi/2$. This algorithm requires 40 ms to reconstruct the refractive index map at one z-position, with a 3-megapixels field of view.

**WPT on standard samples.** To validate the capability of WPT for extracting the RI distribution, we imaged $2\ \mu m$ polystyrene micro-beads (Polysciences Inc.) z-stacks with the RI value of 1.59 at the central wavelength. The beads are suspended in immersion oil (Zeiss) with the RI value of 1.518. Figure 2a shows the three frames corresponding to the different phase shifts of the polystyrene beads. For these experiments, we use 63X/1.4NA objective. The real parts of the correlation function $\Gamma_{is}$ at different time-lapse values are illustrated in Fig. 2b. The RI distribution of the micro-bead for each z slice is reconstructed via Eq. (2). The 3D rendering of the RI distribution described in Fig. 2c was obtained in Amira (Thermo Fisher Scientific). The reconstructed RI value of the microbead agrees well with the expected value of 1.59 at the central wavelength. The halo artifact associated with phase-contrast and SLIM images was removed from the final refractive index maps using our previously reported algorithm [51].

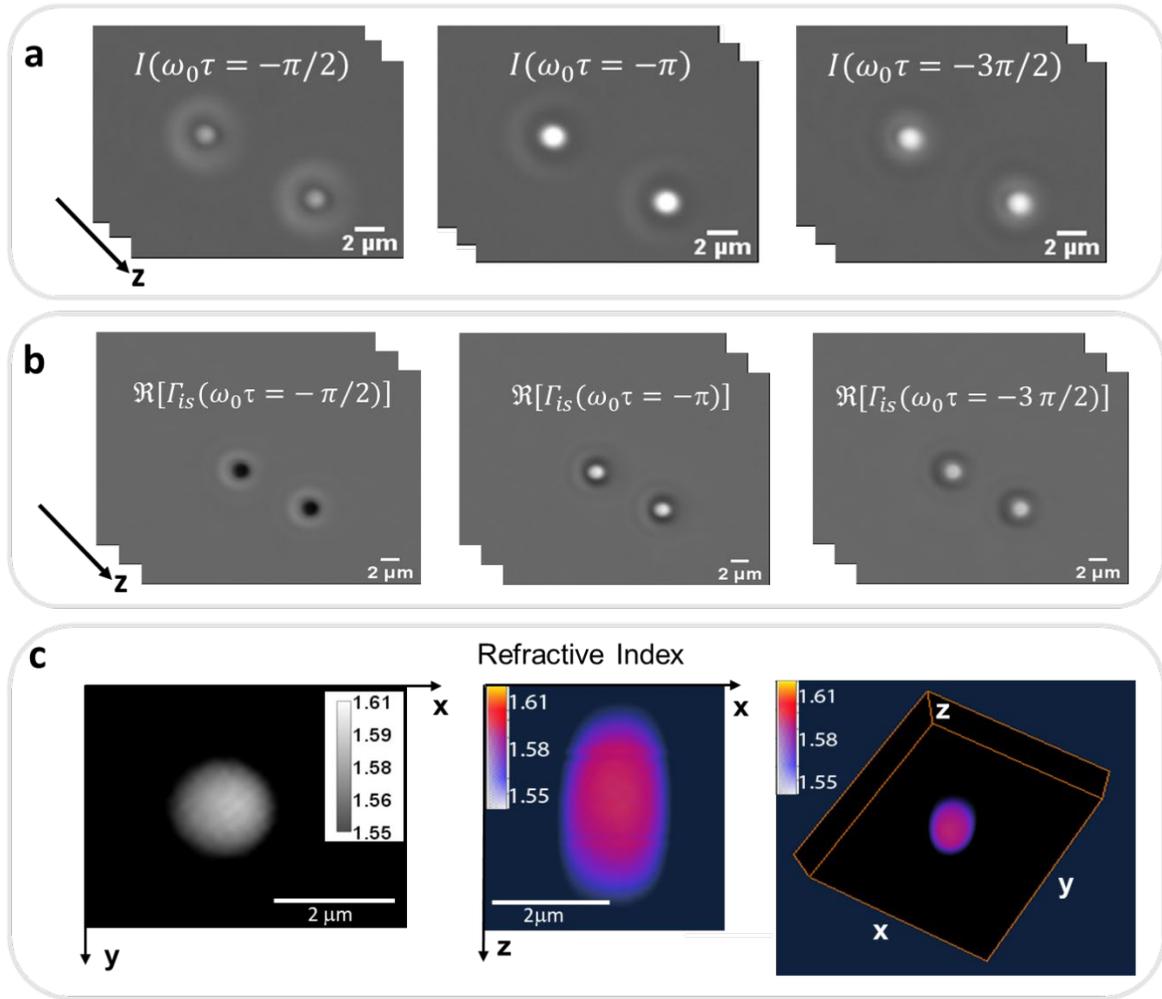

Fig. 2 **WPT on standard samples. a** Three phase-shifted frames of 2μm polystyrene beads suspended in oil are imaged by SLIM using a 63x/1.4NA objective. **b** The real part of the correlation function at three different time-delays is obtained by solving equation (1). **c** 3D RI tomogram of the 2 μm polystyrene bead.

**WPT of sperm cells.** The 3D rendering of a bovine sperm cell is displayed in Fig. 3a (Supplemental video 1). In the sperm head, the acrosome and the nucleus can be identified with RI values between 1.35 and 1.37. The centriole and mitochondria-rich midpiece of the sperm cell yield high refractive index values (Fig. 3b). The tail of the sperm has the RI value of 1.35 and the axial filament inside the tail with a slightly higher RI value of 1.36 can be recognized. The end piece of the sperm has the lowest RI value of approximately 1.34.

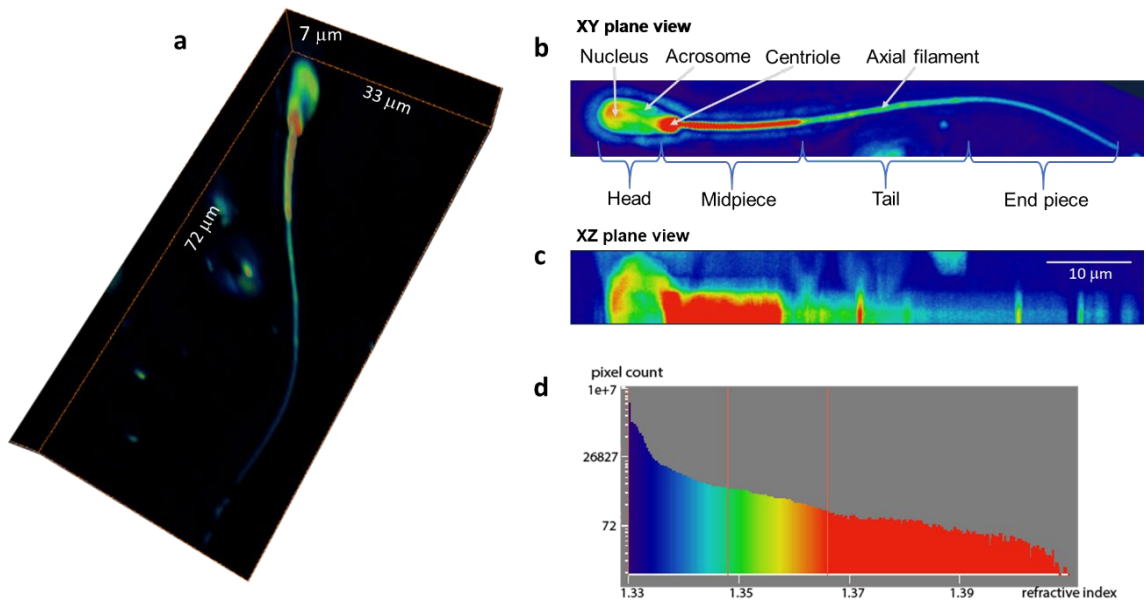

Fig. 3 **WPT of sperm cells. a** 3D RI tomogram of a spermatozoon (40x/0.75NA objective). **b** XY plane projection view. The nucleus, acrosome, centriole, and axial filament of the sperm cell are pointed by the white arrows. **c** XZ plane projection view. **d** Histogram of the RI of the sperm cell.

**WPT of neurons.** Applying the WPT principle, the three frames of hippocampal neurons and their correlation functions are depicted in Figs. 4a-b. The reconstructed RI distribution and 3D rendering of the neuron (Supplemental video 2) are displayed in Figs 4c-d. The more detailed structures of individual hippocampal neurons (Supplemental video 3,4) are illustrated in Figs 4e-f. The rendering in this case used two colormaps, as shown in Figs. 4e and 4f. It can be seen that the neuron dendrites have an RI value of approximately 1.34, while the cell body ranges from 1.35 to 1.38, with a nucleolus of 1.39-1.4. The axon can be recognized in Fig. 4f, as the morphology shows longer and thinner filamentous structure.

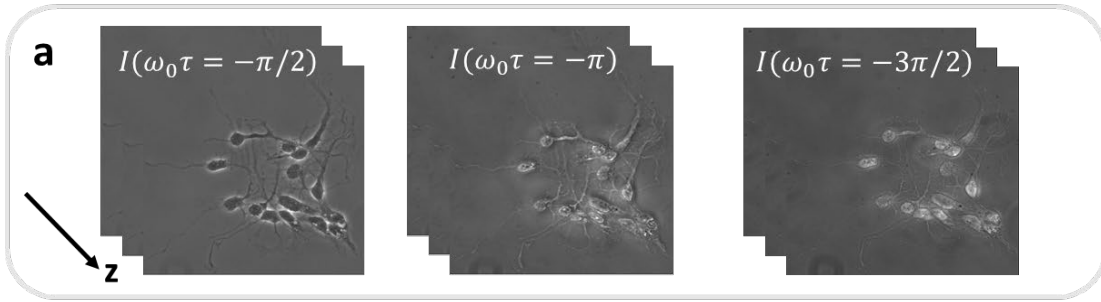

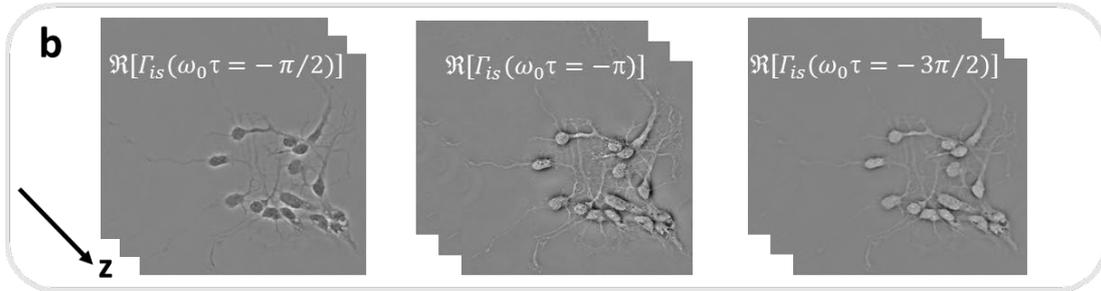

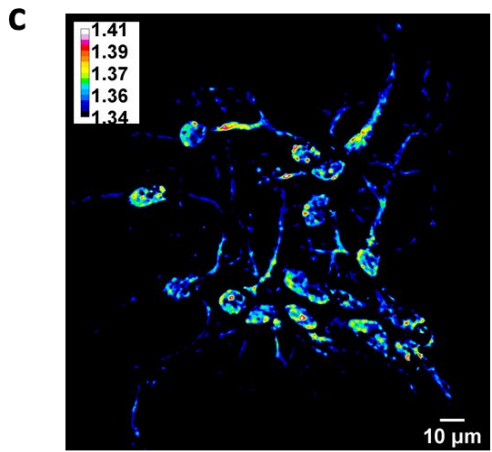

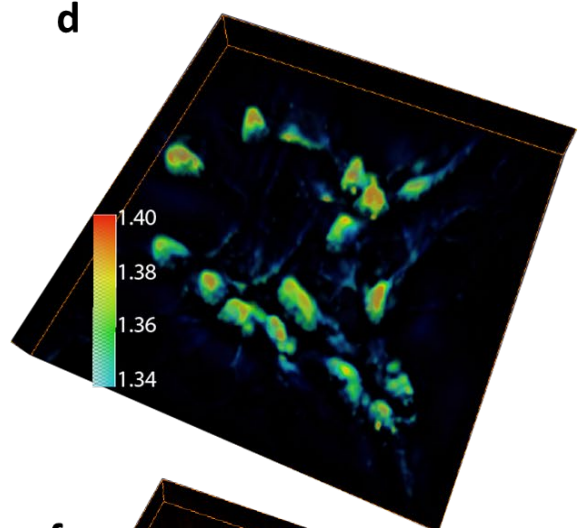

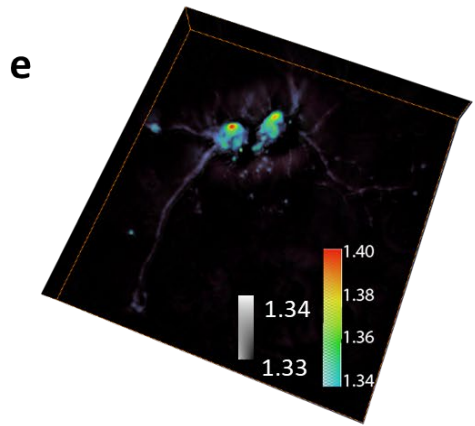

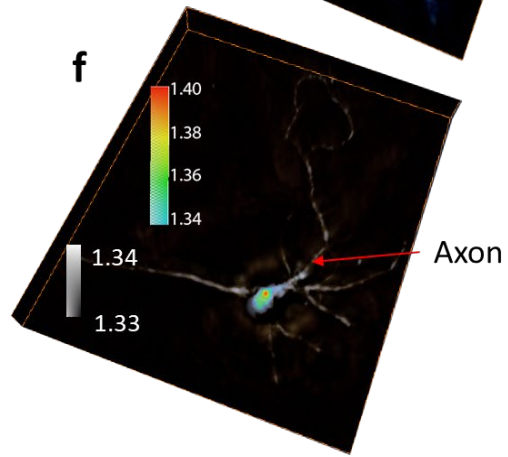

Fig. 4 **WPT of neurons. a** Three phase-shifted frames of hippocampal neurons (40x/0.75NA objective). **b** The real part of the correlation function at three different time-lags are solved from equation (1). **c** RI map of the hippocampal neurons. **d,e,f** 3D rendering of RI tomograms of the hippocampal neurons. (**e,f**) use of two colormaps as indicated to enhance the dendrites and axons. The axon is pointed with a red arrow.

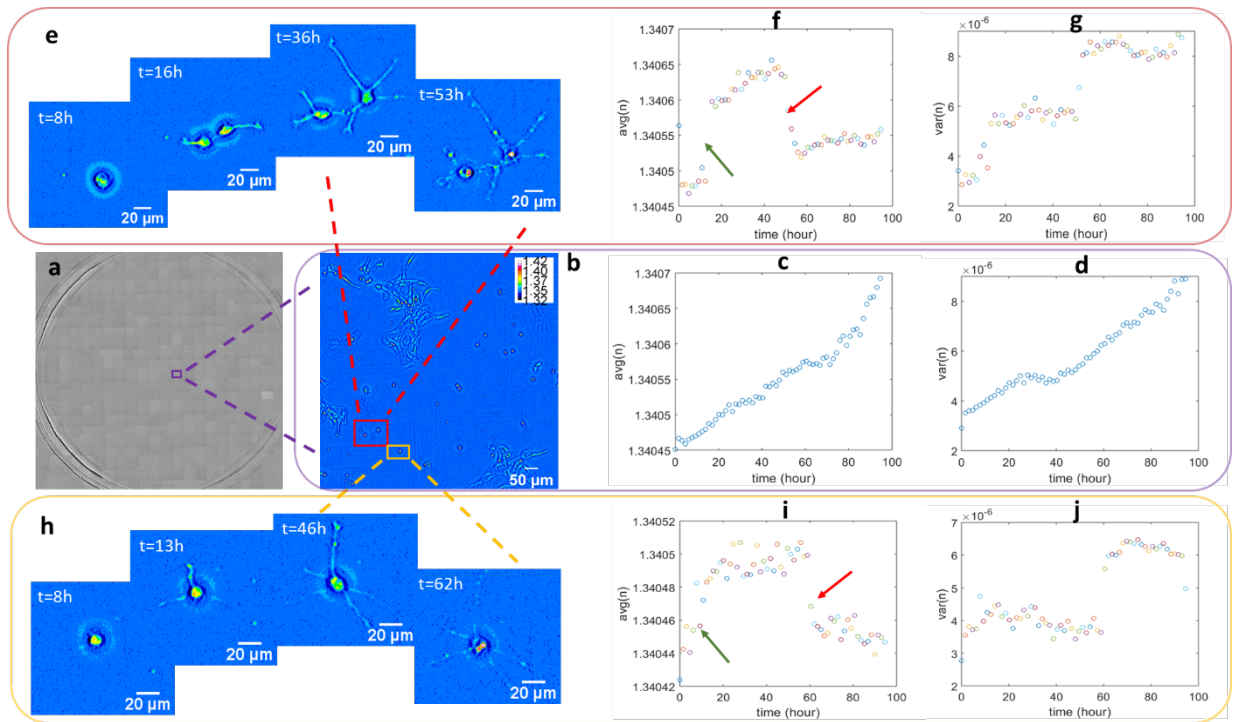

Fig. 5 **Dynamic WPT of live cells across multi-well plates. a** RI map across a whole well of living hippocampal neurons (10x/0.3NA objective) is composed of 20 x 21 mosaic tiles, each of 214 x 204 μm$^2$ area. **b** Zoom-in RI map of the purple box in (**a**) with average (**c**) and variance (**d**) of the RI vs. time. **e** Zoom-in RI map of the red box in (**b**) with average (**f**) and variance (**g**) of the RI. The green arrow indicates the increase of the RI when the two neurons separated and their dendrites appeared, the red arrow shows the decrease in RI when the two neurons died. **h** Zoom-in RI map of the yellow box in (**b**) with average (**i**) and variance (**j**) of the RI. The green arrow indicates the jump of the RI when the dendrites appeared, the red arrow shows the decrease in RI when the neuron died.

**Dynamic WPT of live cells.** Due to the high throughput, low phototoxicity, absence of photobleaching, and easy sample preparation, WPT is capable of studying real-time volumetric biological events in living cells. We imaged the growth and proliferation of hippocampal neurons over the course of several days, in 6-well plates. The RI distribution of the whole well of neurons is displayed in Fig. 5a (Supplemental video 5). One tile zoom-in of the whole well and its distribution of RI are shown in Fig. 5b (Supplemental video 6). Figure 5c describes the average of the RI within this tile versus time. The average RI values increase with time due to neuron growth. Figure 5c illustrates the average RI of the whole tile, including neurons and background. As the neurons grow, more pixels in the region of interest (ROI) appear with higher RI, thus, the average RI becomes larger. Another point worth mentioning is that the range of the y-axis in Fig. 5c is from 1.34045 to 1.34070. Thus, due to the averaging over the large field of view, the change in the RI value detected by our system is on the fifth decimal, indicating the high sensitivity of WPT. Figure 5d shows that the variance of the RI for this tile increases with time as well [52]. Notice that the range of RI variance values is in the order of $10^{-6}$, which is detectable due to the sensitivity conferred by the common path stability and lack of speckles in SLIM.

Figure 5e is the zoom-in image of the red box in Fig. 5b containing two neurons. The neurons spread out into two regions at around t=16 hours, kept growing until around t=53 hours and then died. We can see that both the average and variance of the RI show three different stages (Figs. 5f-g). One significant change in the average and variance of the RI appears when the two neurons separated (red arrows). Another change is clearly visible when the two neurons died (green arrows). The death event was accompanied by a decrease in the mean RI, likely due to the membrane permeability, which allowed for water influx.

Figure 5h is a zoom-in image of the yellow box in Fig. 5b containing one neuron. The neuron dendrites started to appear at approximately t=13 hours time point, resulting in a jump in the average RI (Fig. 5i). The neuron kept growing until approximately t=62 hours and then died, leading to a decrease in the average RI. Some oscillations in the variance (Fig. 5j) of the RI appear before the neuron died, while exhibiting a clear change after the neuron died. Figure 6 demonstrates the capability of WPT for 3D real-time live-cell imaging. The change in the morphology of the neuron can be recognized at different time frames.

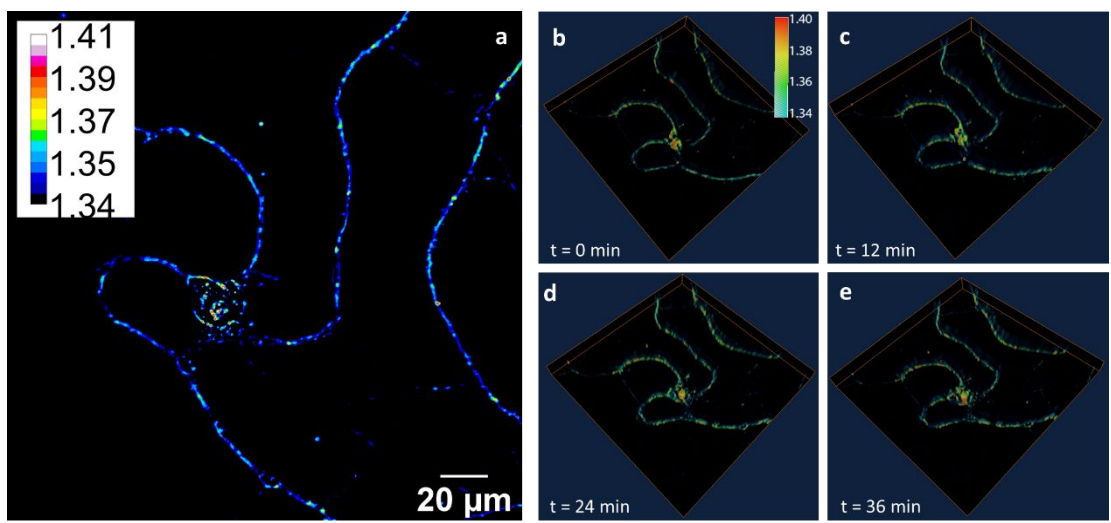

Fig. 6 **Time-lapse WPT of live neurons.** **a** RI map of a live hippocampal neuron imaged with 40x/0.75 NA objective. **b-e** 3D RI tomograms of the hippocampal neuron at 12 mins apart.

## 3  Discussion

In summary, we proposed a new high-throughput RI tomography method, WPT, based on the correlation propagation of partially coherent light. Our model builds on the coherence theory pioneered by Emil Wolf and performs the reconstruction directly in the space-time domain, without the need for Fourier transformation. WPT works with a phase contrast microscope and the SLIM add-on module. The white light illumination and common-path interferometry allow for speckle-free and nanometer path-length stability. High-resolution tomograms of RI distributions are rendered using a z-stack of three phase-shifted intensity frames. WPT decouples the RI distribution from the thickness of the object by calculating the Laplacian and the second-order time derivative of the complex correlation functions. As a result, the tomographic reconstruction is very fast, requiring only 40 ms per z-slice.

We demonstrated the capability of WPT with tomographic reconstructions of standard polystyrene beads, fixed spermatozoa, and hippocampal neurons. WPT has a high RI sensitivity, on the order of $10^{-5}$, which is useful as an intrinsic marker for live cell monitoring. We illustrated this ability by imaging dynamic live cells over many hours. As a label-free method, WPT is nondestructive and not limited by photobleaching and phototoxicity commonly associated with fluorescence microscopy.

We envision that WPT will find important applications in material and life sciences such as studying cell growth with the segmentation of nucleus compared to the whole cell, cell classification using RI, histopathology for cancer diagnosis based on RI [10], and 3D tracking of collagen fibers. However, WPT is still limited by the "missing cone problem" due to the finite numerical aperture. Therefore, WPT can also adopt other methods such as cell-rotating or illumination angle-scanning to achieve better axial resolution and depth sectioning. At the same time, new advances in deep learning look promising in addressing this issue by frequency extrapolation [53].

## 4  Materials and Methods

SLIM add-on module

The SLIM add-on module is mounted to the output camera port of a commercial phase contrast microscope. The SLIM module contains an SLM (Meadowlark) and a camera (Hamamatsu, V2 Orca Flash). The measurements were conducted by using an Axio Observer Z1 microscope (Zeiss) using a halogen light source (Zeiss), with an incubation system, under 63x/1.4 NA, 40x/0.75 NA, 10x/0.3 NA objectives and fully opened condenser at 0.55 NA.

The 40 ms reconstruction per frame is faster than the SLIM image acquisition rate of 250 ms, which requires 30 ms for SLM stabilization (Meadowlark XY Series), 10 ms for exposure (Hamamatsu, V2 Orca Flash), and 90 ms for z-scanning.

Sample preparation

The hippocampal neurons were prepared as follows. Primary hippocampal neurons were harvested from dissected hippocampi of Sprague-Dawley rat embryos. Hippocampi were dissociated with the enzyme in order to have hippocampal neurons. Hippocampal neurons were then plated on to a six-well plate that is pre-coated with poly-D-lysine (0.1 mg/ml; Sigma-Aldrich). Hippocampal neurons were initially incubated with a plating medium containing 86.55% MEM Eagle's with Earle's BSS (Lonza), 10% Fetal Bovine Serum (re-filtered, heat-inactivated; Thermo Fisher), 0.45% of 20% (wt./vol.) glucose, 1x 100 mM sodium pyruvate (100x; Sigma-Aldrich), 1x 200 mM glutamine (100x; Sigma-Aldrich), and 1x Penicillin/ Streptomycin (100x; Sigma-Aldrich) in order to help attachment of neurons(300 cells/mm2). After three hours of incubation in an incubator (37°C and 5% CO2), the plating media was aspirated and replaced with maintenance media containing Neurobasal growth medium supplemented with B-27 (Invitrogen), 1% 200 mM glutamine (Invitrogen) and 1% penicillin/streptomycin (Invitrogen) at 37 °C, in the presence of 5% CO2. The hippocampal neurons in Fig. 5 were grown for 2 days in vitro and took dynamic images for 4 days. The hippocampal neurons in Fig. 6 were grown for 14 days in vitro and took snapshots every 12 mins. The hippocampal neurons in Fig. 4 were fixed. The sperm cell in Fig. 3 was fixed in 10% paraformaldehyde.

**Data availability statement:**

The data that support the findings of this study are available from the corresponding author upon reasonable request.

**Availability of computer code and algorithms:**

The code and computer algorithms that support the findings of this study are available from the corresponding author upon reasonable request.

**Contributions:**

G.P. and X.C. proposed the idea. X.C. solved the inverse problem and derived the model. M.E.K. wrote the control software. X.C., M.E.K., and C.H. performed imaging. X.C. wrote the reconstruction algorithm and analyzed the data. Y.L. prepared the hippocampal neurons. X.C., M.E.K., and G.P. wrote the manuscript. G.P. supervised the work.

**Acknowledgments:**

This work is supported by National Science Foundation (CBET0939511 STC, NRT-UtB 173525), National Institute of General Medical Sciences (GM129709); National Cancer Institute (CA238191).

**Conflict of Interest:**

The authors declare the following competing interests: G.P. has a financial interest in Phi Optics, Inc., a company developing quantitative phase imaging technology for materials and life science applications.

Supplementary information for

**Wolf phase tomography (WPT) of transparent structures using partially coherent illumination**

*Xi Chen, Mikhail E. Kandel, Chenfei Hu, Young Jae Lee, and Gabriel Popescu*

**Supplementary Note 1: WPT reconstruction model**

The stochastic scalar field generated by a partially coherent source propagating in a media with refractive index distribution $n(\mathbf{r})$ with scattering potential defined to be $\chi(\mathbf{r}) = n^2(\mathbf{r}) - n_0^2$ is the average of the refractive index of the media. Each realization of the stochastic incident field satisfies Helmholtz equation [1]

$$\nabla_2^2 U_i(\mathbf{r}_2, \omega) + k_0^2 n_0^2 U_i(\mathbf{r}_2, \omega) = 0, \tag{1}$$

where $k_0 = \omega/c$ is the wavenumber in vacuum, the $\nabla_2^2$ is the Laplacian operator taken with respect to the point $\mathbf{r}_2$. While every realization of the total field $U(\mathbf{r}) = U_i(\mathbf{r}) + U_s(\mathbf{r})$ obeys the inhomogeneous Helmholtz equation

$$\nabla_2^2 U(\mathbf{r}_2,\omega) + k_0^2 n^2(\mathbf{r}_2) U(\mathbf{r}_2,\omega) = 0. \tag{2}$$

Subtracting Eq. (1) from Eq. (2), we arrive at

$$\nabla_2^2 U_s(\mathbf{r}_2,\omega) + k_0^2 \chi(\mathbf{r}_2) U_i(\mathbf{r}_2,\omega) + k_0^2 n^2(\mathbf{r}_2) U_s(\mathbf{r}_2,\omega) = 0. \tag{3}$$

Multiplying both sides of Eq. (3) by $U_i^*(\mathbf{r}_1,\omega)$, we have

$$\nabla_2^2 [U_i^*(\mathbf{r}_1,\omega) U_s(\mathbf{r}_2,\omega)] + k_0^2 \chi(\mathbf{r}_2) U_i^*(\mathbf{r}_1,\omega) U_i(\mathbf{r}_2,\omega)$$

$$+ k_0^2 n^2(\mathbf{r}_2) U_i^*(\mathbf{r}_1,\omega) U_s(\mathbf{r}_2,\omega) = 0. \tag{4}$$

Note that we may place $U_i^*(\mathbf{r}_1,\omega)$ under the Laplacian operator because that it was taken with respect to $\mathbf{r}_2$. Taking the ensemble average of all the realizations will return the correlation propagation function

$$\nabla_2^2 W_{is}(\mathbf{r}_1,\mathbf{r}_2,\omega) + k_0^2 \chi(\mathbf{r}_2) W_{ii}(\mathbf{r}_1,\mathbf{r}_2,\omega) + k_0^2 n^2(\mathbf{r}_2) W_{is}(\mathbf{r}_1,\mathbf{r}_2,\omega) = 0, \tag{5}$$

where the spectral density function is defined as

$$W_{pq}(\mathbf{r}_1,\mathbf{r}_2,\omega) = \langle U_p^*(\mathbf{r}_1,\omega) U_q(\mathbf{r}_2,\omega) \rangle, \qquad p,q = \{i,s\}. \tag{6}$$

Be aware that Eq. (5) is not true anymore if $\mathbf{r}_1 = \mathbf{r}_2$ due to the fact that the Laplacian does not only take on one position. Instead, two more terms will appear in the expression

$$\nabla^2 W_{is}(\mathbf{r},\mathbf{r},\omega) - 2\langle \nabla U_i^*(\mathbf{r},\omega) \cdot \nabla U_s(\mathbf{r},\omega) \rangle + k_0^2 n_0^2 W_{is}(\mathbf{r},\mathbf{r},\omega)$$

$$+ k_0^2 \chi(\mathbf{r}) W_{ii}(\mathbf{r},\mathbf{r},\omega) + k_0^2 n^2(\mathbf{r}) W_{is}(\mathbf{r},\mathbf{r},\omega) = 0, \tag{7}$$

Compared to other terms, the contribution from the second term is negligible. Since the source is broadband and partially coherent, we can express it as a summation of plane waves propagating in different directions

$$U_i(\mathbf{r},\omega) = \sum_{j=1}^{\infty} A_j(\omega) e^{i(\mathbf{k}_{0j}\cdot\mathbf{r}+\psi_j)}. \tag{8}$$

The scattered field with Born approximation and far-zone approximation has the expression

$$\begin{aligned}U_s(\mathbf{r},\omega) &= \frac{e^{ik_0 r}}{r}\int_V F(\mathbf{r}')U_i(\mathbf{r}')e^{-ik_0\frac{\mathbf{r}}{|\mathbf{r}|}\cdot\mathbf{r}'}dr'^3 \\ &= \frac{e^{ik_0 r}}{r}\int_V F(\mathbf{r}')\sum_{j=1}^{\infty} A_j(\omega)e^{i(\mathbf{k}_{0j}\cdot\mathbf{r}+\phi_j)}e^{-ik_0\frac{\mathbf{r}}{|\mathbf{r}|}\cdot\mathbf{r}'}d^3r',\end{aligned} \tag{9}$$

Where $F = \frac{k_0^2}{4\pi}[n^2(\mathbf{r})-1]$ is the scattering potential. Now let us consider the term $\langle \nabla U_i^*(\mathbf{r},\omega)\cdot\nabla U_s(\mathbf{r},\omega)\rangle$. Because the incident field is almost uniform in the field of view, thus the derivatives along x and y direction are approximately zero, only the derivative along z-axis contributes to the gradient of the incident field. While the scattered field is a modulated spherical wave determined by the integral in Eq. (9). The lateral resolution is smaller than the axial resolution. It implies that the gradient of the scattered field is mostly along x and y directions, the z component is approximately zero. Therefore, the gradient of the incident field is approximately perpendicular to the gradient of the scattered field, resulting in the ensemble average of the second term in Eq. (7) is negligible. Dropping off the second term, we finally have

$$\nabla^2 W_{is}(\mathbf{r},\mathbf{r},\omega) + k_0^2[n_0^2 + n^2(\mathbf{r})]W_{is}(\mathbf{r},\mathbf{r},\omega) + k_0^2 \chi(\mathbf{r})W_{ii}(\mathbf{r},\mathbf{r},\omega) = 0. \qquad (10)$$

Mutual coherence function relates to the cross-spectral function via Fourier transform

$$\Gamma(\mathbf{r},\mathbf{r},\tau) = \int_0^\infty W(\mathbf{r},\omega)e^{-i\omega\tau}d\omega. \qquad (11)$$

Taking the Fourier transformation on both sides of Eq. (10), it becomes

$$\nabla^2 \Gamma_{is}(\mathbf{r},\mathbf{r},\tau) - \frac{[n_0^2 + n^2(\mathbf{r})]}{c^2}\frac{\partial^2 \Gamma_{is}(\mathbf{r},\mathbf{r},\tau)}{\partial \tau^2} - \frac{\chi(\mathbf{r})}{c^2}\frac{\partial^2 \Gamma_{ii}(\mathbf{r},\mathbf{r},\tau)}{\partial \tau^2} = 0. \qquad (12)$$

Now if we express the complex mutual coherence function as $\Gamma_{is}(\mathbf{r},\mathbf{r},\tau) = \Re[\Gamma_{is}(\mathbf{r},\mathbf{r},\tau)] + i\Im[\Gamma_{is}(\mathbf{r},\mathbf{r},\tau)]$, where $\Re$ and $\Im$ denote the real and imaginary parts. Taking the real part and evaluating at $\omega_0\tau = -\pi$, Eq. (12) becomes

$$\nabla^2 \Re[\Gamma_{is}(\mathbf{r},\mathbf{r},-\pi)] - \frac{[n_0^2 + n^2(\mathbf{r})]}{c^2}\frac{\partial^2 \Re[\Gamma_{is}(\mathbf{r},\mathbf{r},\omega_0\tau)]}{\partial \tau^2}\bigg|_{\omega_0\tau=-\pi} - \frac{\chi(\mathbf{r})}{c^2}\frac{\partial^2 \Re[\Gamma_{ii}(\mathbf{r},\mathbf{r},\omega_0\tau)]}{\partial \tau^2}\bigg|_{\omega_0\tau=-\pi} = 0.$$

(13)

Therefore, the refractive index distribution of the media is

$$n(\mathbf{r}) = \sqrt{\frac{m - n_0^2(1-g)}{1+g}}, \qquad (14)$$

Where

$$m(\mathbf{r}) = \frac{c^2 \nabla^2 \Re[\Gamma_{is}(\mathbf{r},\mathbf{r},\omega_0\tau)]}{\dfrac{\partial^2 \Re[\Gamma_{is}(\mathbf{r},\mathbf{r},\omega_0\tau)]}{\partial \tau^2}}\bigg|_{\omega_0\tau=-\pi},$$

$$g(\mathbf{r}) = \frac{\left.\frac{\partial^2 \Re[\Gamma_{ii}(\mathbf{r},\mathbf{r},\omega_0\tau)]}{\partial \tau^2}\right|}{\left.\frac{\partial^2 \Re[\Gamma_{is}(\mathbf{r},\mathbf{r},\omega_0\tau)]}{\partial \tau^2}\right|_{\omega_0\tau=-\pi}}.$$

The real part of the mutual correlation function is readily obtained with the SLIM add-on interferometer. Instead of recovering the phase information, as in typical processing, we use three phase-contrast frames to recover the refractive index as follows. The three frames in SLIM are

$$I_d(\mathbf{r}) = I_i(\mathbf{r}) + I_s(\mathbf{r}) + 2\Re\left[\Gamma_{is}\left(\omega_0\tau_d + \Delta\phi(\mathbf{r})\right)\right], \tag{15}$$

where $\omega_0\tau_d = -d\pi/2$, $d = 1,2,3$, $\omega_0$ is the central frequency of the incident field, $\Re$ stands for the real part, $\Delta\phi$ is the phase difference between the incident field, $U_i$, and scattered field, $U_s$, and $\Gamma_{pq}(\mathbf{r}_1,\mathbf{r}_2,\tau) = \langle U_p^*(\mathbf{r}_1,t)U_q(\mathbf{r}_2,t+\tau)\rangle_t$, $p,q = \{i,s\}$. From these three frames, we are able to solve for $\Re\left[\Gamma_{is}\left(\omega_0\tau_d + \Delta\phi(\mathbf{r})\right)\right]$. In the experiment, we used partially coherent annular source, which can be characterized by the partially coherent Schell model, meaning the degree of coherence is only a function of the difference between two different positions on the source [2]. The cross-spectral function of Schell model has the form

$$W_{ii}(\mathbf{r}_1,\mathbf{r}_2,\omega) = \iint p(\mathbf{v})a^*(\mathbf{r}_1,\omega)e^{2\pi i\mathbf{r}_1\cdot\mathbf{v}}a(\mathbf{r}_2,\omega)e^{-2\pi i\mathbf{r}_2\cdot\mathbf{v}}d\mathbf{v}$$

$$= a^*(\mathbf{r}_1,\omega)a(\mathbf{r}_2,\omega)\iint p(\mathbf{v})e^{-2\pi i\mathbf{r}_d\cdot\mathbf{v}}d\mathbf{v}, \tag{16}$$

where $\mathbf{r}_d \equiv \mathbf{r}_2 - \mathbf{r}_1$, $a(\mathbf{r},\omega)$ is the amplitude of the incident field at position $\mathbf{r}$, $p(\mathbf{v})$ is the spectral density of the incident field in the source plane. Thus $p(\mathbf{v})$ is constant when $R_1 \leq |\mathbf{v}| \leq R_2$ and zero otherwise, where $R_1$ and $R_2$ are the inner and outer radii of the phase ring in the condenser. After the integration over the source plane, Eq. (16) becomes

$$W_{ii}(\mathbf{r}_1,\mathbf{r}_2,\omega) = a^*(\mathbf{r}_1,\omega)a(\mathbf{r}_2,\omega)\frac{R_2 J_1(2\pi|\mathbf{r}_d|R_2) - R_1 J_1(2\pi|\mathbf{r}_d|R_1)}{\pi(R_2^2 - R_1^2)|\mathbf{r}_d|}, \qquad (17)$$

where $J_1$ is Bessel function of the first kind. At the same argument, it becomes the spectral density

$$S(\mathbf{r},\omega) = W(\mathbf{r},\mathbf{r},\omega) = |a(\mathbf{r},\omega)|^2. \qquad (18)$$

Therefore, the mutual coherence function of the incident field is

$$\Gamma_{ii}(\mathbf{r},\tau) = \int_0^\infty |a(\mathbf{r},\omega)|^2 e^{-i\omega\tau} d\omega. \qquad (19)$$

Supplemental video 1: 3D RI rendering of the sperm cell

Supplemental video 2: 3D RI rendering of the neurons

Supplemental video 3: 3D RI rendering of the individual neurons

Supplemental video 4: 3D RI rendering of the neuron with dendrites and the axon

Supplemental video 5: 2D RI zoom-out of the whole well of the neuron

Supplemental video 6: Time-lapse of 2D RI of the neurons